\documentclass[aps,prl,twocolumn,groupedaddress]{revtex4}
\usepackage{graphicx}

\begin{document}

\title{The Anderson Transition in Two-Dimensional Systems with Spin-Orbit Coupling}

\author{Yoichi Asada and Keith Slevin}
\email[]{slevin@phys.sci.osaka-u.ac.jp}
\affiliation{Department of Physics, Graduate School of Science, 
Osaka University, 1-1 Machikaneyama, Toyonaka, Osaka 560-0043, Japan}

\author{Tomi Ohtsuki}
\affiliation{Department of Physics, 
Sophia University, Kioi-cho 7-1, Chiyoda-ku, Tokyo 102-8554, Japan}

\date{\today}

\begin{abstract}
We report a numerical investigation of the Anderson transition in
two-dimensional systems with spin-orbit coupling.
An accurate estimate of the critical exponent $\nu$ for the divergence
of the localization length in this universality
class has to our knowledge not been reported in the literature.
Here we analyse the SU(2) model. We find that for this
model corrections to scaling due to irrelevant scaling variables may
be neglected permitting an accurate estimate of the
exponent $\nu=2.73 \pm 0.02$.
\end{abstract}


\maketitle

According to the scaling theory of localisation \cite{abrahams:79},
if interactions are neglected, all states are localised in two
dimensions (2D).
Two exceptions predicted by the scaling theory to this oft recited 
mantra are the extended states which occur at the center of a 
Landau level in the quantum Hall effect (QHE) \cite{huckestein:95},
and the Anderson transition which occurs in zero magnetic field if 
there is a significant spin-orbit interaction \cite{hikami:80,ando:89}.

A surprising aspect of the QHE is that the estimate
$\nu=2.35\pm.03$ \cite{huckestein:90,huckestein:95} for 
non-interacting electrons of the 
critical exponent $\nu$, which describes the divergence
of the localisation length $\xi$ at the transition,  
is in close agreement with the measured value. 
This is despite the fact that interactions are clearly relevant
since the dynamical exponent $z$ is predicted to be two for 
non-interacting electrons while the experimental value is unity.

Critical phenomena are determined by the symmetry of the Hamiltonian 
and the dimensionality of the system.
The important symmetries for the Anderson transition exhibited
by non-interacting electrons are time reversal symmetry and spin 
rotation symmetry.
There are three universality classes: orthogonal, unitary and symplectic.
Systems with time reversal symmetry but where spin rotation
symmetry is broken by the spin-orbit interaction
belong to the symplectic class.

In this paper we estimate using numerical simulation and
finite size scaling the exponent $\nu$ for the Anderson transition 
in the symplectic universality class in 2D.
In contrast to the QHE, where extended states occur only in
vanishingly small energy region, the metallic phase extends
over a finite energy interval.  
This system is, therefore, a good candidate for the study of 2D 
quantum phase transitions.
Early work suggested that metallic phase in this model is
destroyed when interactions between electrons are taken into 
account \cite{altshuler:83,castellani:84b},
while more recent work suggests that this is not so \cite{castellani:98}.

Recently, there have been numerous reports of the observation of 
a zero magnetic field metal-insulator transition in 2D together with 
measurements of the critical exponents which characterize this transition
\cite{abrahams:01}.
Whether or not those experiments indicate the existence of a 
true metallic phase at zero temperature remains in dispute.
If there is indeed a transition, the physics driving it and whether
there is a relation with the transition we study here is not
yet clear. However, just as was the case for the QHE, we believe that an 
accurate estimate for the metal-insulator transition in zero field 
in non-interacting 2D systems may prove useful.

\begin{table}
\caption{\label{table:nu}
Published estimates of the critical exponent for the
2D symplectic universality class. Q1DLL means finite 
size scaling for the quasi-one dimensional localization length, ELS energy level statistics
and MFSS multifractal finite size scaling. The errors quoted are one
standard deviation.}
\begin{ruledtabular}
\begin{tabular}{llll}
Ref. & model & method & $\nu$ \\ \hline
\onlinecite{ando:89}       & Ando model      & Q1DLL & $2.05 \pm 0.08$ \\
\onlinecite{fastenrath:91} & Ando model      & Q1DLL & $2.75 \pm 0.1$  \\
\onlinecite{evangelou:95}  & Evangelou model & Q1DLL & $2.5  \pm 0.3$  \\
\onlinecite{schweitzer:97} & Ando model      & ELS   & $2.32 \pm 0.14$ \\
\onlinecite{merkt:98}      & network model   & Q1DLL & $2.51 \pm 0.18$ \\
\onlinecite{minakuchi:98}  & network model   & Q1DLL & $2.88 \pm 0.15$ \\
\onlinecite{yakubo:98}     & Ando model      & MFSS & $2.41 \pm 0.24$ \\
\end{tabular}
\end{ruledtabular}
\end{table}

There has been only very limited success in estimating the
critical exponent with field theoretic methods \cite{hikami:92,brezin:97}.
In Table \ref{table:nu} we tabulate the estimates of the
exponent reported in previous numerical studies
\cite{ando:89,fastenrath:91,evangelou:95,schweitzer:97,merkt:98,
minakuchi:98,yakubo:98}.
There is considerable variation between these estimates. The 
estimates reported in \cite{fastenrath:91,evangelou:95,merkt:98,
minakuchi:98,yakubo:98} seem to be consistent with a true value of
the exponent in the range $\left[ 2.6,2.9 \right]$.
However, the estimate of \cite{schweitzer:97} is somewhat below
this and that of \cite{ando:89} is in contradiction with the estimates
of \cite{fastenrath:91,minakuchi:98}. 

We use the transfer matrix method 
\cite{pichard:81,mackinnon:83} to estimate the localization length $\lambda$ 
of electrons on an $L \times L_z$ quasi-1D strip and
then extract the critical exponent from a finite size scaling 
analysis of the dependence of $\lambda$ on $L$ and disorder.
Two important factors limiting the accuracy of the estimate of the exponent
obtained in this way are the accuracy of the data for $\lambda$
and the maximum width $L$ for which data are available.
The standard error in the estimate of $\lambda$ decreases as 
$\sqrt{\lambda/L_z}$, while for a fixed $L_z$ the CPU time needed
increases as $L^3$.
At the critical point $\lambda$ increases linearly with $L$ so that
the CPU time needed to estimate $\lambda$ to a given accuracy 
increases as $L^4$.
This means that it is somewhat easier to improve the accuracy of the
numerical data than to increase the size of the systems simulated.
However, as the accuracy of the raw data improves, corrections to scaling
due to irrelevant scaling variables become more important.
While such corrections can be taken into account \cite{slevin:99}, the number of
fitting parameters is increased and correspondingly the uncertainty
in the estimate of the exponent is increased.
It is therefore advantageous to choose a model for which such corrections are
negligible even when the raw data are of high accuracy. 
In this paper we report results for an SU(2) model for which 
this condition is satisfied.

The Hamiltonian for the SU(2) model describes non-interacting electrons
on a 2D square lattice with nearest neighbour hopping
\begin{equation}
H=\sum_{i,\sigma} \epsilon_i c_{i,\sigma}^{\dagger}c_{i,\sigma}
- V \sum_{<i,j>,\sigma,\sigma'} R(i,j)_{\sigma \sigma'}
c_{i,\sigma}^{\dagger}c_{j,\sigma'}
\label{hamiltonian}
\end{equation}
Here $c^{\dagger}_{i,\sigma}$ ($c_{i,\sigma}$)
denotes the creation (annihilation)
operator of an electron at the site $i$ with spin $\sigma$ and
$\epsilon_i$ denotes the random potential at site $i$.
We assume a box distribution
with each $\epsilon_i$ uniformly and independently distributed on the interval
$[-W/2,W/2]$.
The width $W$ of the distribution measures the strength of the randomness.
The constant $V$ is taken to be the unit of energy, $V=1$.

The spin-orbit coupling appears in the hopping matrix $R(i,j)$ between 
each pair of nearest neighbours on the lattice.
These matrices belong to the group SU(2) of 2 $\times$ 2 unitary
matrices with determinant one.
The hopping  matrices are parameterised as follows
\begin{eqnarray}
R(i,j)
\!
=
\! \!
\left(
\! \!
\begin{array}{cc}
e^{i\alpha_{i,j}} \cos \beta_{i,j}
& e^{i\gamma_{i,j}} \sin \beta_{i,j} \\
-e^{-i\gamma_{i,j}} \sin \beta_{i,j}
& e^{-i\alpha_{i,j} }\cos \beta_{i,j}
\end{array}
\! \!
\right) \label{hoppingmatrix}
\! \!
\end{eqnarray}
This matrix describes a rotation of the electron spin
in three dimensional space. (The Euler angles of this 
rotation are related to, but not equal to, the angles $\alpha$,
$\beta$ and $\gamma$.)
In the SU(2) model the distribution of these angles
is chosen so that the $R(i,j)$ are uniformly
distributed with respect to the group invariant 
measure (Haar measure) on SU(2). 
This corresponds to $\alpha$ and $\gamma$ uniformly
distributed in the range $[0,2\pi)$,
and $\beta$ distributed according to the probability density,
\begin{equation}
P(\beta) \rm d \beta = 
\left\{
\begin{array}{ll}
\sin (2\beta) \rm d \beta & 0 \le \beta \le \frac{\pi}{2} \\ 
 0                                  & \rm otherwise.
\end{array}
\right.
\end{equation}
Hopping matrices on different bonds of the lattice are statistically
independent.
Periodic boundary conditions are imposed in the transverse direction.
The necessary calculations are carried out using quaternion
arithmetic \cite{ando:89} which halves the
required number of multiplications compared with an
implementation using complex arithmetic.

Some of the physics of the SU(2) model can be understood by
comparing it with the Ando model which has been adopted 
in \cite{ando:89,fastenrath:91,schweitzer:97,yakubo:98}. 
In the Ando model as the electron propagates through the material
its spin precesses at a rate and about an axis which depend on the
electrons wave number. When scattered by the random
potential, the rate and the axis about which the electron's spin 
rotates changes. 
This leads to a diffusive motion of the spin
with an associated spin relaxation length. Quantum interference 
between time reversed electron trajectories longer than this
length produces the weak anti-localization effect \cite{bergmann:82}. 
Motivated by the conjecture that the spin relaxation length might be
an important irrelevant length scale, we adopted the SU(2) model 
where the uniform distribution
of the hopping matrices on SU(2) ensures that the spin relaxation
length is the shortest possible.
Doing so we do indeed find that corrections due to irrelevant scaling
variables can be neglected.

To determine the critical exponent $\nu$, critical disorder $W_c$ and 
other critical properties
of the transition we fit dependence of $\Lambda=\lambda/L$ on
the system size $L$ and the disorder $W$, or when $W=0$ on the Fermi energy $E_F$,
to a one parameter scaling law of the form
\begin{equation}
\ln \Lambda = F \left(\psi  \right).
\label{a1}
\end{equation}
Here $\psi$ is the relevant scaling variable.
We expand the scaling function as a power series
\begin{equation}
F\left(x \right) = \ln \Lambda_c + x + a_2 x^2 + \ldots
+ a_{n_0}x^{n_0}
\label{a2}
\end{equation}
terminating the expansions at order $n_0$.
To allow for non-linearity of the scaling variable, the
scaling variable is approximated by an expansion in terms of the 
dimensionless disorder 
$w=(W_c-W)/W_c$, where $W_c$ is the critical
disorder separating the insulating and metallic phases.
(If $W=0$ we set $w=(E_F-E_c)/E_c$.)
The growth of the relevant scaling variable with system size is described
by the critical exponent $\nu$
\begin{equation}
\psi = L^{1/\nu} \left(\psi_1 w + \psi_2 w^2 + \ldots + \psi_{n_{\psi}} w^{n_{\psi}} \right) ,
\label{a3}
\end{equation}
where we terminate the expansion at order $n_{\psi}$.
This same exponent describes the divergence of the localization (correlation) length
\begin{equation}
\xi = \xi_{\pm} \left|\psi_1 w + \psi_2 w^2 + \ldots + \psi_{n_{\psi}} w^{n_{\psi}} \right|^{-\nu}.
\end{equation}
where we terminate the expansion at order $n_{\psi}$.
The absolute scales $\xi_{\pm}$ of the localization length on either side
of the transition
are not determined in this analysis, so we set them both to unity for
simplicity.
The linear coefficient in the expansions of $F$ is set to unity, 
as shown, to eliminate some redundancy in the definition of the fitting
parameters. 
The total number of parameters is $N_p = n_0 + n_{\psi} +2$.

The best fit is determined by minimizing the ${\chi}^2$ statistic.
The quality of the fit is assessed with the goodness of fit 
probability $Q$.
Confidence intervals for the fitted parameters are estimated
using a Monte Carlo method \cite{numrep}. This involves using the model and the best
estimates of the fitting parameters to generate an ideal data set.
From this data set a large ensemble of synthetic data sets is
generated by adding random errors, with a variance equal to
that of the error of the corresponding data point, to the ideal data set.
Fitting of the ensemble of synthetic data sets produces a
distribution for the critical parameters from which confidence intervals
and the goodness of fit are estimated. This procedure
is standard and systematic but does not take into account
any unknown systematic effects that might only be discernible for
very much larger systems.
Of course, this caveat applies to almost any numerical estimate of 
a critical exponent.

\begin{figure}
\includegraphics{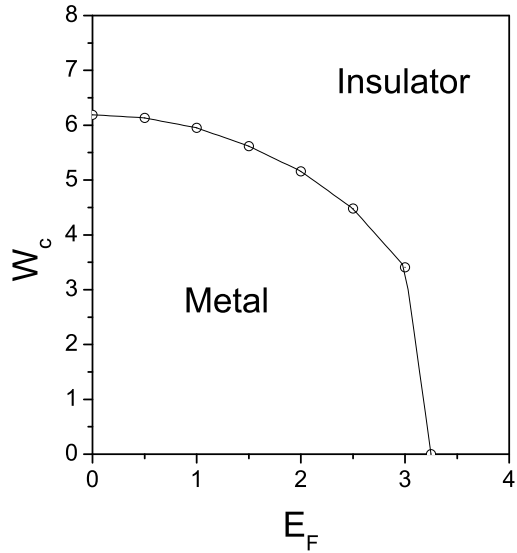}
\caption{\label{fig:pd}
The phase diagram for the SU(2) model.}
\end{figure}

Before turning to the estimate of the critical exponent
we sketch the phase diagram in Figure \ref{fig:pd}.
The figure is based on data for systems with sizes ranging from $L=8$ to 
$L=32$ for a number of Fermi energies between 
$E_F=0$ and $E_F=3$ for which the critical disorder $W_c$ was estimated,
and on data for $W=0$ for which the critical energy $E_c$ was estimated.
In the absence of a random potential i.e. when $W=0$, the Hamiltonian may 
have chiral symmetry and, in addition to the transition at finite $E_c$, 
a critical state may also be present at the band center \cite{brouwer:00}. 
Whether this possibility is realised depends on the boundary conditions 
and whether the number of sites is even or odd.
Since chiral symmetry is broken by a random potential it does not
affect our estimation of $\nu$ below.

\begin{figure}
\includegraphics{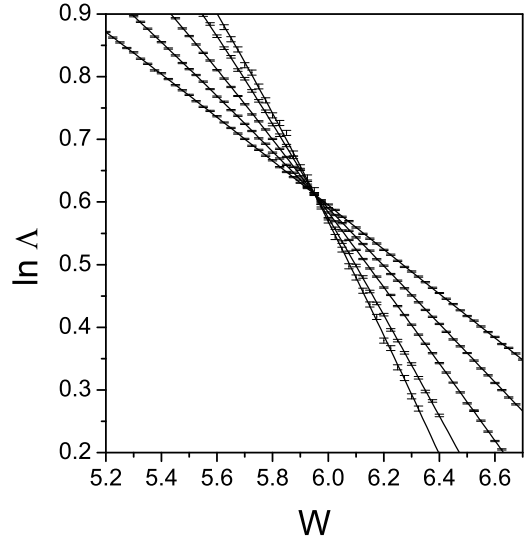}
\caption{\label{fig:su2a}
The numerical data for the SU(2) model and the best fit. 
Data for $L=8,16,32,64$ and $96$ are shown.}
\end{figure}

\begin{table}
\caption{\label{table:su2a}
The details of various fits to the numerical data for the SU(2) model.
The fit is to $N_d=230$ data points in the range $5.2 \le W \le 6.7$
and $0.2 \le \ln \Lambda \le 0.9$.}
\begin{ruledtabular}
\begin{tabular}{llllllll}
$n_0$  & $n_{\psi}$  & $N_p$ &  $Q$ & $W_c$ & $\ln \Lambda_c$ & $\nu$
\\ \hline
2  &  2  & 6 &   0.2   & 5.952$\pm$ .002 & 0.612$\pm$ .001   & 2.74$\pm$ .01  \\
3  &  2  & 7 &   0.4   & 5.952$\pm$ .002 & 0.612$\pm$ .001   & 2.73$\pm$ .02  \\
4  &  2  & 8 &   0.3   & 5.952$\pm$ .002 & 0.612$\pm$ .001   & 2.73$\pm$ .02  \\
3  &  3  & 8 &   0.3   & 5.952$\pm$ .002 & 0.612$\pm$ .001   & 2.74$\pm$ .03  \\
4  &  3  & 9 &   0.3   & 5.952$\pm$ .002 & 0.612$\pm$ .001   & 2.73$\pm$ .03  \\
\end{tabular}
\end{ruledtabular}
\end{table}

\begin{table}
\caption{\label{table:su2b}
The variation of the estimates of the critical parameters for the SU(2) model as data
for smaller systems sizes are progressively excluded from consideration.
Here $n_0=3$ and $n_{\psi}=2$. }
\begin{ruledtabular}
\begin{tabular}{llllllll}
  & $N_d$   &  $Q$ & $W_c$ & $\ln \Lambda_c$ & $\nu$
\\ \hline
$8 \le L \le 96$  &  230   &   0.4   & 5.952$\pm$ .002 & 0.612$\pm$ .001   & 2.73$\pm$ .02  \\
$16\le L \le 96$  &  169   &   0.6   & 5.953$\pm$ .003 & 0.611$\pm$ .002   & 2.75$\pm$ .02  \\
$32\le L \le 96$  &  113   &   0.5   & 5.954$\pm$ .005 & 0.611$\pm$ .003   & 2.71$\pm$ .04  \\
$64\le L \le 96$  &  64    &   0.8   & 5.96$\pm$ .02 & 0.60$\pm$ .02   & 2.8$\pm$ .2  \\
\end{tabular}
\end{ruledtabular}
\end{table}

\begin{table}
\caption{\label{table:su2c}
The variation of the estimates of the critical parameters for the SU(2) model
as the range of disorder under consideration is progressively narrowed.}
\begin{ruledtabular}
\begin{tabular}{llllllll}
  W & $N_d$  & $n_0$ & $n_{\psi}$  & $Q$ & $W_c$ & $\ln \Lambda_c$ & $\nu$ \\ \hline
$\left[ 5.2, 6.7 \right]$  &  230   &   3 & 2  & 0.4 & 5.952$\pm$ .002 & 0.612$\pm$ .001   & 2.73$\pm$ .02  \\
$\left[ 5.5 ,6.4 \right]$  &  175   &   2 & 2 & 0.7 & 5.953$\pm$ .003 & 0.612$\pm$ .001   & 2.75$\pm$ .02  \\
$\left[ 5.8 , 6.1 \right]$  &  65    &   1 & 1  & 0.5 & 5.950$\pm$ .002 & 0.613$\pm$ .002   & 2.72$\pm$ .07  \\
\end{tabular}
\end{ruledtabular}
\end{table}

\begin{figure}
\includegraphics{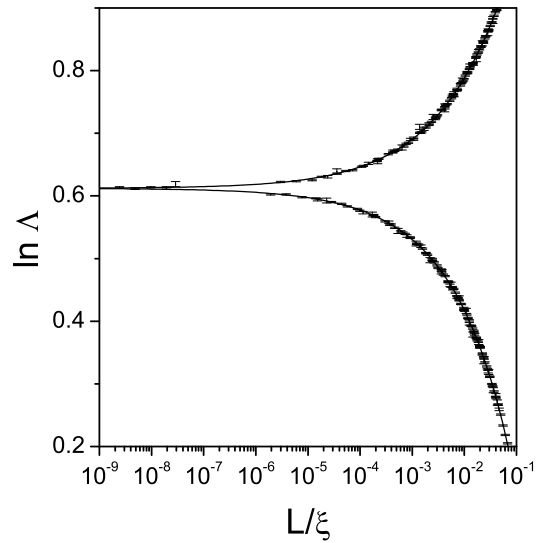}
\caption{\label{fig:su2b}
When plotted versus $L/\xi$ the data divide naturally into an upper
metallic branch $F_+\left(L/\xi \right)$ and a lower insulating branch
$F_-\left(L/\xi \right)$.}
\end{figure}

To estimate the critical exponent accurately more
extensive simulations were carried out for a single energy
$E_F=1$.
The numerical data are presented in Figures \ref{fig:su2a} and \ref{fig:su2b}.
Data with an accuracy of $0.1\%$ are available for system sizes $L=8,16$
and $32$, with accuracy $0.2\%$ for $L=64$, and $0.4\%$ for $L=96$.
This required $L_z$ of the order of $10^7$ to $10^8$ depending on 
the size, the disorder and the accuracy.
When fitting the data the intervals of $W$ and $\ln \Lambda$ 
to consider must be decided. The exact choice is not particularly
important provided all data are in the critical regime.
A larger interval of $\ln \Lambda$ requires a higher order of
expansion in Eq. \ref{a2}, while a larger interval of $W$
requires a higher order expansion in Eq. \ref{a3}.
The results of the finite size analysis are presented in Tables \ref{table:su2a},
\ref{table:su2b} and \ref{table:su2c}.
A number of fits of the numerical data are possible but, as
can be seen by referring to Table \ref{table:su2a}, all yield
consistent results.
The estimates of the critical parameters are also stable against
restriction of the system sizes under consideration, see Table \ref{table:su2b},
and against a narrowing of the range of disorder, see Table \ref{table:su2c}.
The lines shown in Figure \ref{fig:su2a} and \ref{fig:su2b} correspond to
a fit with $n_0=3$ and $n_{\psi}=2$. 
To demonstrate single parameter scaling graphically we
re-plot the data as a function of $L/\xi$ in Figure \ref{fig:su2b};
metallic and insulating branches are clearly visible.
These are described by related 
scaling functions $F_+$ and $F_-$ where
\begin{equation}
F_{\pm}\left(x \right)= \ln \Lambda_c \pm x^{1/\nu} + a_2 x^{2/\nu} + \ldots
+ \left(\pm1 \right)^{n_0}a_{n_0}x^{n_0/\nu}
\end{equation}

In summary, we have studied the Anderson transition in the 2D symplectic 
universality class and estimated the critical exponent $\nu$.
We find $\nu=2.73\pm .02$,
where the error is a $95\%$ confidence interval. Our result is consistent
with the estimates reported in \cite{fastenrath:91,evangelou:95,merkt:98,
minakuchi:98,yakubo:98} but not with those of 
\cite{ando:89,schweitzer:97}. Analyses based on energy level statistics,
such as \cite{schweitzer:97},
also seem to have a tendency to produce estimates which are lower 
than those of the transfer matrix method for the 3D orthogonal
universality class \cite{romer:02}.
On the other hand, in our opinion, the error bar for the exponent claimed in
\cite{ando:89} is too optimistic.

\bibliography{su2}

\begin{thebibliography}{24}
\expandafter\ifx\csname natexlab\endcsname\relax\def\natexlab#1{#1}\fi
\expandafter\ifx\csname bibnamefont\endcsname\relax
  \def\bibnamefont#1{#1}\fi
\expandafter\ifx\csname bibfnamefont\endcsname\relax
  \def\bibfnamefont#1{#1}\fi
\expandafter\ifx\csname citenamefont\endcsname\relax
  \def\citenamefont#1{#1}\fi
\expandafter\ifx\csname url\endcsname\relax
  \def\url#1{\texttt{#1}}\fi
\expandafter\ifx\csname urlprefix\endcsname\relax\def\urlprefix{URL }\fi
\providecommand{\bibinfo}[2]{#2}
\providecommand{\eprint}[2][]{\url{#2}}

\bibitem[{\citenamefont{Abrahams et~al.}(1979)\citenamefont{Abrahams, Anderson,
  Licciardello, and Ramakrishnan}}]{abrahams:79}
\bibinfo{author}{\bibfnamefont{E.}~\bibnamefont{Abrahams}},
  \bibinfo{author}{\bibfnamefont{P.~W.} \bibnamefont{Anderson}},
  \bibinfo{author}{\bibfnamefont{D.~C.} \bibnamefont{Licciardello}},
  \bibnamefont{and} \bibinfo{author}{\bibfnamefont{T.~V.}
  \bibnamefont{Ramakrishnan}}, \bibinfo{journal}{Phys. Rev. Lett.}
  \textbf{\bibinfo{volume}{42}}, \bibinfo{pages}{673} (\bibinfo{year}{1979}).

\bibitem[{\citenamefont{Huckestein}(1995)}]{huckestein:95}
\bibinfo{author}{\bibfnamefont{B.}~\bibnamefont{Huckestein}},
  \bibinfo{journal}{Rev. Mod. Phys.} \textbf{\bibinfo{volume}{67}},
  \bibinfo{pages}{357} (\bibinfo{year}{1995}).

\bibitem[{\citenamefont{Hikami et~al.}(1980)\citenamefont{Hikami, Larkin, and
  Nagaoka}}]{hikami:80}
\bibinfo{author}{\bibfnamefont{S.}~\bibnamefont{Hikami}},
  \bibinfo{author}{\bibfnamefont{A.~I.} \bibnamefont{Larkin}},
  \bibnamefont{and} \bibinfo{author}{\bibfnamefont{Y.}~\bibnamefont{Nagaoka}},
  \bibinfo{journal}{Prog. Theor. Phys.} \textbf{\bibinfo{volume}{63}},
  \bibinfo{pages}{707} (\bibinfo{year}{1980}).

\bibitem[{\citenamefont{Ando}(1989)}]{ando:89}
\bibinfo{author}{\bibfnamefont{T.}~\bibnamefont{Ando}}, \bibinfo{journal}{Phys.
  Rev. B} \textbf{\bibinfo{volume}{40}}, \bibinfo{pages}{5325}
  (\bibinfo{year}{1989}).

\bibitem[{\citenamefont{Huckestein and Kramer}(1990)}]{huckestein:90}
\bibinfo{author}{\bibfnamefont{B.}~\bibnamefont{Huckestein}} \bibnamefont{and}
  \bibinfo{author}{\bibfnamefont{B.}~\bibnamefont{Kramer}},
  \bibinfo{journal}{Phys. Rev. Lett.} \textbf{\bibinfo{volume}{64}},
  \bibinfo{pages}{1437} (\bibinfo{year}{1990}).

\bibitem[{\citenamefont{Altshuler and Aronov}(1983)}]{altshuler:83}
\bibinfo{author}{\bibfnamefont{B.~L.} \bibnamefont{Altshuler}}
  \bibnamefont{and} \bibinfo{author}{\bibfnamefont{A.~G.}
  \bibnamefont{Aronov}}, \bibinfo{journal}{Solid State Commun.}
  \textbf{\bibinfo{volume}{46}}, \bibinfo{pages}{429} (\bibinfo{year}{1983}).

\bibitem[{\citenamefont{Castellani et~al.}(1984)\citenamefont{Castellani,
  Castro, Forgacs, and Sorella}}]{castellani:84b}
\bibinfo{author}{\bibfnamefont{C.}~\bibnamefont{Castellani}},
  \bibinfo{author}{\bibfnamefont{C.~D.} \bibnamefont{Castro}},
  \bibinfo{author}{\bibfnamefont{G.}~\bibnamefont{Forgacs}}, \bibnamefont{and}
  \bibinfo{author}{\bibfnamefont{S.}~\bibnamefont{Sorella}},
  \bibinfo{journal}{Solid State Commun.} \textbf{\bibinfo{volume}{52}},
  \bibinfo{pages}{261} (\bibinfo{year}{1984}).

\bibitem[{\citenamefont{Castellani et~al.}(1998)\citenamefont{Castellani,
  Castro, and Lee}}]{castellani:98}
\bibinfo{author}{\bibfnamefont{C.}~\bibnamefont{Castellani}},
  \bibinfo{author}{\bibfnamefont{C.~D.} \bibnamefont{Castro}},
  \bibnamefont{and} \bibinfo{author}{\bibfnamefont{P.~A.} \bibnamefont{Lee}},
  \bibinfo{journal}{Phys. Rev. B} \textbf{\bibinfo{volume}{57}},
  \bibinfo{pages}{R9381} (\bibinfo{year}{1998}).

\bibitem[{\citenamefont{Abrahams et~al.}(2001)\citenamefont{Abrahams,
  Kravchenko, and Sarachik}}]{abrahams:01}
\bibinfo{author}{\bibfnamefont{E.}~\bibnamefont{Abrahams}},
  \bibinfo{author}{\bibfnamefont{S.~V.} \bibnamefont{Kravchenko}},
  \bibnamefont{and} \bibinfo{author}{\bibfnamefont{M.~P.}
  \bibnamefont{Sarachik}}, \bibinfo{journal}{Rev. Mod. Phys.}
  \textbf{\bibinfo{volume}{73}}, \bibinfo{pages}{251} (\bibinfo{year}{2001}).

\bibitem[{\citenamefont{Fastenrath et~al.}(1991)\citenamefont{Fastenrath,
  Adams, Bundschuh, Hermes, Raab, Schlosser, Wehner, and
  Wichmann}}]{fastenrath:91}
\bibinfo{author}{\bibfnamefont{U.}~\bibnamefont{Fastenrath}},
  \bibinfo{author}{\bibfnamefont{G.}~\bibnamefont{Adams}},
  \bibinfo{author}{\bibfnamefont{R.}~\bibnamefont{Bundschuh}},
  \bibinfo{author}{\bibfnamefont{T.}~\bibnamefont{Hermes}},
  \bibinfo{author}{\bibfnamefont{B.}~\bibnamefont{Raab}},
  \bibinfo{author}{\bibfnamefont{I.}~\bibnamefont{Schlosser}},
  \bibinfo{author}{\bibfnamefont{T.}~\bibnamefont{Wehner}}, \bibnamefont{and}
  \bibinfo{author}{\bibfnamefont{T.}~\bibnamefont{Wichmann}},
  \bibinfo{journal}{Physica A} \textbf{\bibinfo{volume}{172}},
  \bibinfo{pages}{302} (\bibinfo{year}{1991}).

\bibitem[{\citenamefont{Evangelou}(1995)}]{evangelou:95}
\bibinfo{author}{\bibfnamefont{S.~N.} \bibnamefont{Evangelou}},
  \bibinfo{journal}{Phys. Rev. Lett.} \textbf{\bibinfo{volume}{75}},
  \bibinfo{pages}{2550} (\bibinfo{year}{1995}).

\bibitem[{\citenamefont{Schweitzer and Zharekeshev}(1997)}]{schweitzer:97}
\bibinfo{author}{\bibfnamefont{L.}~\bibnamefont{Schweitzer}} \bibnamefont{and}
  \bibinfo{author}{\bibfnamefont{I.~K.} \bibnamefont{Zharekeshev}},
  \bibinfo{journal}{J. Phys. Cond. Matt.} \textbf{\bibinfo{volume}{9}},
  \bibinfo{pages}{L441} (\bibinfo{year}{1997}).

\bibitem[{\citenamefont{Merkt et~al.}(1998)\citenamefont{Merkt, Janssen, and
  Huckestein}}]{merkt:98}
\bibinfo{author}{\bibfnamefont{R.}~\bibnamefont{Merkt}},
  \bibinfo{author}{\bibfnamefont{M.}~\bibnamefont{Janssen}}, \bibnamefont{and}
  \bibinfo{author}{\bibfnamefont{B.}~\bibnamefont{Huckestein}},
  \bibinfo{journal}{Phys. Rev. B} \textbf{\bibinfo{volume}{58}},
  \bibinfo{pages}{4394} (\bibinfo{year}{1998}).

\bibitem[{\citenamefont{Minakuchi}(1998)}]{minakuchi:98}
\bibinfo{author}{\bibfnamefont{K.}~\bibnamefont{Minakuchi}},
  \bibinfo{journal}{Phys. Rev. B} \textbf{\bibinfo{volume}{58}},
  \bibinfo{pages}{9627} (\bibinfo{year}{1998}).

\bibitem[{\citenamefont{Yakubo and Ono}(1998)}]{yakubo:98}
\bibinfo{author}{\bibfnamefont{K.}~\bibnamefont{Yakubo}} \bibnamefont{and}
  \bibinfo{author}{\bibfnamefont{M.}~\bibnamefont{Ono}},
  \bibinfo{journal}{Phys. Rev. B} \textbf{\bibinfo{volume}{58}},
  \bibinfo{pages}{9767} (\bibinfo{year}{1998}).

\bibitem[{\citenamefont{Hikami}(1992)}]{hikami:92}
\bibinfo{author}{\bibfnamefont{S.}~\bibnamefont{Hikami}},
  \bibinfo{journal}{Prog. Theor. Phys. Suppl.} \textbf{\bibinfo{volume}{107}},
  \bibinfo{pages}{213} (\bibinfo{year}{1992}).

\bibitem[{\citenamefont{Brezin and Hikami}(1997)}]{brezin:97}
\bibinfo{author}{\bibfnamefont{E.}~\bibnamefont{Brezin}} \bibnamefont{and}
  \bibinfo{author}{\bibfnamefont{S.}~\bibnamefont{Hikami}},
  \bibinfo{journal}{Phys. Rev. B} \textbf{\bibinfo{volume}{55}},
  \bibinfo{pages}{R10169} (\bibinfo{year}{1997}).

\bibitem[{\citenamefont{Pichard and Sarma}(1981)}]{pichard:81}
\bibinfo{author}{\bibfnamefont{J.-L.} \bibnamefont{Pichard}} \bibnamefont{and}
  \bibinfo{author}{\bibfnamefont{G.}~\bibnamefont{Sarma}}, \bibinfo{journal}{J.
  Phys. C} \textbf{\bibinfo{volume}{14}}, \bibinfo{pages}{L127}
  (\bibinfo{year}{1981}).

\bibitem[{\citenamefont{MacKinnon and Kramer}(1983)}]{mackinnon:83}
\bibinfo{author}{\bibfnamefont{A.}~\bibnamefont{MacKinnon}} \bibnamefont{and}
  \bibinfo{author}{\bibfnamefont{B.}~\bibnamefont{Kramer}},
  \bibinfo{journal}{Z. Phys. B} \textbf{\bibinfo{volume}{53}},
  \bibinfo{pages}{1} (\bibinfo{year}{1983}).

\bibitem[{\citenamefont{Slevin and Ohtsuki}(1999)}]{slevin:99}
\bibinfo{author}{\bibfnamefont{K.}~\bibnamefont{Slevin}} \bibnamefont{and}
  \bibinfo{author}{\bibfnamefont{T.}~\bibnamefont{Ohtsuki}},
  \bibinfo{journal}{Phys. Rev. Lett.} \textbf{\bibinfo{volume}{82}},
  \bibinfo{pages}{382} (\bibinfo{year}{1999}).

\bibitem[{\citenamefont{Bergmann}(1982)}]{bergmann:82}
\bibinfo{author}{\bibfnamefont{G.}~\bibnamefont{Bergmann}},
  \bibinfo{journal}{Solid State Commun.} \textbf{\bibinfo{volume}{42}},
  \bibinfo{pages}{815} (\bibinfo{year}{1982}).

\bibitem[{\citenamefont{Press et~al.}(1992)\citenamefont{Press, Teukolsky,
  Vetterling, and Flannery}}]{numrep}
\bibinfo{author}{\bibfnamefont{W.~H.} \bibnamefont{Press}},
  \bibinfo{author}{\bibfnamefont{A.~A.} \bibnamefont{Teukolsky}},
  \bibinfo{author}{\bibfnamefont{W.~T.} \bibnamefont{Vetterling}},
  \bibnamefont{and} \bibinfo{author}{\bibfnamefont{B.~P.}
  \bibnamefont{Flannery}}, \emph{\bibinfo{title}{Numerical Recipes in Fortran}}
  (\bibinfo{publisher}{Cambridge University Press},
  \bibinfo{address}{Cambridge}, \bibinfo{year}{1992}).

\bibitem[{\citenamefont{Brouwer et~al.}(2000)\citenamefont{Brouwer, Mudry, and
  Furusaki}}]{brouwer:00}
\bibinfo{author}{\bibfnamefont{P.}~\bibnamefont{Brouwer}},
  \bibinfo{author}{\bibfnamefont{C.}~\bibnamefont{Mudry}}, \bibnamefont{and}
  \bibinfo{author}{\bibfnamefont{A.}~\bibnamefont{Furusaki}},
  \bibinfo{journal}{Phys. Rev. Lett.} \textbf{\bibinfo{volume}{84}},
  \bibinfo{pages}{2913} (\bibinfo{year}{2000}).

\bibitem[{\citenamefont{Ndawana et~al.}(2002)\citenamefont{Ndawana,
  {R\"{o}mer}, and Schreiber}}]{romer:02}
\bibinfo{author}{\bibfnamefont{M.~L.} \bibnamefont{Ndawana}},
  \bibinfo{author}{\bibfnamefont{R.~A.} \bibnamefont{{R\"{o}mer}}},
  \bibnamefont{and}
  \bibinfo{author}{\bibfnamefont{M.}~\bibnamefont{Schreiber}},
  \bibinfo{journal}{Eur. Phys. J. B} \textbf{\bibinfo{volume}{27}},
  \bibinfo{pages}{399} (\bibinfo{year}{2002}).

\end{thebibliography}

\end{document}